\documentclass[aps,prl,twocolumn,amsmath,amssymb,showpacs]{revtex4}
\usepackage{graphicx}% Include figure files
\usepackage{dcolumn}% Align table columns on decimal point
\usepackage{bm}% bold math
\usepackage{psfrag} % pour modifier des donnees d'un ps
\usepackage{color}
\usepackage{here}
\begin{document}
\title{Phase separation and flux quantization in the doped quantum dimer model on the square 
and triangular lattices}
\author{Arnaud Ralko,${^1}$
Fr\'{e}d\'{e}ric Mila,${^2}$ and Didier Poilblanc${^1}$}
\affiliation{ ${^1}$ Laboratoire de Physique Th\'eorique, CNRS \&
Universit\'e Paul
Sabatier, F-31062 Toulouse, France \\
${^2}$ Institute of Theoretical Physics, Ecole Polytechnique F\'{e}d\'{e}rale
de Lausanne (EPFL), CH-1015 Lausanne, Switzerland
}
\date{April 5, 2007}
\begin{abstract}
The doped two-dimensional quantum dimer model is
investigated by numerical techniques on the square and triangular
lattices, with significantly different results. On the square lattice,
at small enough doping, 
there is always a phase separation between an insulating
valence-bond solid and a uniform superfluid phase, 
whereas on the triangular lattice, doping leads
directly to a uniform superfluid in a large portion of the RVB phase.
Under an applied Aharonov-Bohm flux, the superfluid exhibits
quantization in terms of half-flux quanta, consistent with $Q=2e$
elementary charge quanta in transport properties.
\end{abstract}

\pacs{75.10.Jm, 05.50.+q, 05.30.-d}
\maketitle
\begin{figure}[h]
\vspace{0.5cm}
%\psfrag{RVB}[tc][tc][0.9][0]{~~~~~RVB}
%\psfrag{a}[bc][br][0.9][0]{columnar}
%\psfrag{c}[bc][bc][0.9][0]{staggered}
%\psfrag{l}[bc][bc][0.9][0]{liquid}
%\psfrag{v}[bc][bc][0.9][0]{$\frac{v}{J}$}
%\psfrag{r}[bc][bc][0.7][0]{RK}
%\psfrag{square}[bc][bc][0.9][0]{square}
%\psfrag{triangular}[bc][bc][0.9][0]{triangular}
%\psfrag{bt}[bc][bc][0.9][0]{$\sqrt{12} \times \sqrt{12}$}
%\psfrag{bs}[bc][bc][0.9][0]{plaquette}
\includegraphics[width=0.45\textwidth,clip]{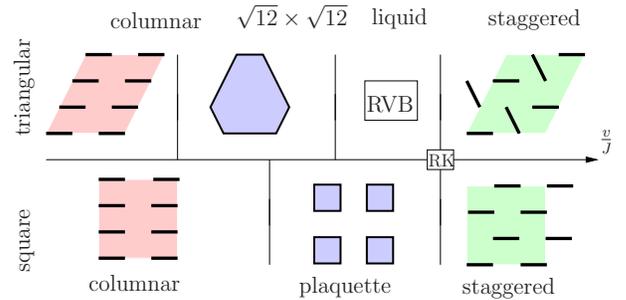}
\caption{\label{fig01}  (color online) Schematic phase diagrams for the
triangular and the square lattice.}
\end{figure}

% }}}1
%------------------------------------------------------------------------------
% Paper
%------------------------------------------------------------------------------
% text {{{1
Understanding electron pairing in high temperature superconductors
is a major challenge in strongly correlated systems. In his
milestone paper, Anderson proposed a simple connection between
high temperature superconductors and Mott
insulators \cite{anderson}. Electron pairs "hidden" in the strongly
correlated insulating parent state as Valence Bond (VB) singlets
lead, once fried to move at finite doping, to a superconducting
behavior. A very good candidate of the insulating parent state is
the resonating VB state (RVB), a state with only exponentially
decaying correlations and no lattice symmetry breaking. A simple
realization of RVB has been proposed by Rokhsar and Kivelson (RK)
in the framework of an effective quantum dimer model (QDM) with
only local processes and orthogonal dimer coverings \cite{rokhsar}.
Even though the relevance of these models for the description of
SU$(2)$ Heisenberg models is still debated, this approach
is expected to capture the physics of systems that naturally
possess singlet ground states (GS). For instance, specific quantum dimer
models have recently been derived from a spin-orbital model
describing LiNiO$_2$ \cite{vernay}, or from the trimerized {\it kagome}
antiferromagnet \cite{zhitomirsky}. In a recent work, a family of
doped QDMs (at T=0) generalizing the so-called RK point of Ref.\cite{rokhsar} 
has been constructed and investigated\cite{poilblanc},
taking advantage of a mapping to classical dimer models \cite{Castelnovo} that 
extends the mapping of the RK model onto a classical model at infinite temperature,
with evidence of phase separation at low doping.  
However, the soluble models of Ref.[\onlinecite{poilblanc}] 
are 'ad hoc' constructions,
and this call for the investigation of similars issue in the context
of more realistic models. In that respect, a natural minimal model to describe
the motion of charge carriers in a sea of dimers is the 
two-dimensional quantum hard-core dimer-gas Hamiltonian:
\begin{eqnarray}\label{hamilt}
H &=& v \sum_{c} N_c | c \rangle \langle c | -J \sum_{(c,c')} | c' \rangle
\langle c | -t \sum_{(c,c'')} | c'' \rangle \langle c | \nonumber \\
\end{eqnarray}
where the sum on $(c)$ runs over all configurations of the Hilbert
space, $N_c$ is the number of flippable plaquettes, the sum on $(c',c)$ runs over all
configurations $| c \rangle$ and $| c' \rangle$ that differ by a single
plaquette dimer flip, and
the sum on $(c'',c)$ runs over all configurations $| c \rangle$ and $|
c''\rangle$ that differ by a single hole hopping between nearest neighbors (triangular)
or (diagonal) next-nearest neighbors (square). 
Throughout the energy scale is set by $J=1$. A schematic phase diagram
for the two lattices is depicted  in Fig.\ref{fig01} in the undoped case. Remarkably,
these lattices lead to quite different insulating states. Indeed,
an ordered plaquette phase appears on the square lattice
immediately away from the special RK point, whereas a RVB liquid
phase is present in the triangular lattice
% Figure Phase diagrams {{{2
% }}}2

In this Letter, we investigate in details the properties of model
(\ref{hamilt}) on the square and triangular lattices at finite doping. 
Building on the differences between the two lattices in the undoped case,
we investigate to which extent the properties of
the doped system are governed by the nature of the insulating
parent state. This investigation is based on exact Diagonalisations and
extensive Green's Function Monte-Carlo (GFMC) simulations \cite{nandini}
essentially free of the usual finite-size 
limitations~\cite{ralko}.

\begin{figure}[h]
\vspace{0.0cm}
%\psfrag{st}[tc][tl][1.1][0]{$s_t(x)$}
%\psfrag{ss}[tc][tr][1.1][0]{$s_s(x)$}
%\psfrag{x}[tc][tc][1.0][0]{$x$}
%\psfrag{a}{(a)}
%\psfrag{b}{(b)}
%\psfrag{aa}[tc][tl][1.0][0]{Triangular}
%\psfrag{ab}[tc][tl][0.9][0]{v=0.85, t=0.05}
%\psfrag{ba}[tc][tl][1.0][0]{Square}
%\psfrag{bb}[tc][tl][0.9][0]{v=0.90, t=0.10}
\includegraphics[width=0.45\textwidth,clip]{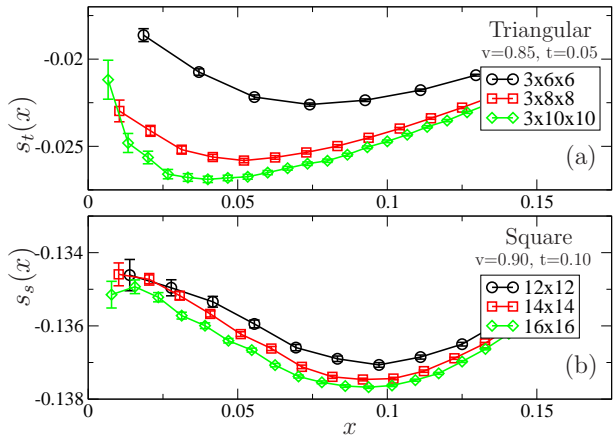}
\caption{\label{fig02}  (color online) Slope of energy density (Eq.(2)) vs
doping for different sizes. (a) Triangular lattice. (b) Square lattice.}
\end{figure}

{\it Phase separation:}
At small $t$, it is expected that holes experience an effective attractive
potential. It is therefore natural to first address the issue of phase
separation (PS), {\it i.e.} the possibility for the system to spontaneously
undergo a macroscopic segregation into two phases with different hole
concentrations. We analyze the problem as a function of the hopping
parameter $t$ and
hole concentration $x = n_h / N$, where $n_h$ is the number
of holes in the system and $N$ the number of sites. In
order to perform a Maxwell construction we define:
\begin{eqnarray}
s(x) = \frac{e(x) - e(0)}{x}
\end{eqnarray}
where $e(x)$ is the energy per site at doping $x$. This quantity corresponds
to the slope of the line passing through $e(0)$ and $e(x)$. If
the
system exhibits PS, the energy will present a change of
curvature implying $s(x)$ to have a minimum at a critical doping $x_c$.
The fact that the local curvature of $e(x)$ at $x=0$ is negative then
implies that the
two seperated phases will have $x=x_c$ and $x=0$ (the undoped
insulator).
In
Fig.\ref{fig02}, typical results are shown for both square and triangular
lattices and for different sizes.
% Figure Maxwell constructions {{{2
% }}}2
Interestingly, PS appears in both cases, but with noticeable differences.
While for the square
lattice (lower panel) the critical hole concentration $x_c$ is
roughly size independent, there is a strong size dependence for the 
triangular lattice (upper panel). This
size effect can be traced back to the nature of the parent undoped state.
On the square lattice, the crystalline phase (for $v<1$) at zero
doping is very robust and for increasing size, its
local order changes only weakly. On the
triangular lattice, it has been shown that size
effects play an important role \cite{ralko}, especially in the RVB liquid phase
for $0.8 \lesssim v \leq 1$. Periodic boundary conditions (BC) tend to stabilize the
so-called $\sqrt{12} \times \sqrt{12}$ phase on small clusters, and 
clusters with more than $192$ sites are necessary to significantly
reduce finite-size effects,
in particular, as in Fig.\ref{fig02}, close to the transition point
with the crystalline phase. Hence the PS
observed around $x=0.075$ for the $3\times 6 \times 6$ cluster is not representative of the
thermodynamic limit.

\begin{figure}[h]
\vspace{0.5cm}
%\psfrag{xc}[bc][tc][1.1][0]{$x_c$}
%\psfrag{n}[bc][bc][1.1][0]{$N^{-1}$}
%\psfrag{t}[tc][tc][1.0][0]{Triangular}
%\psfrag{s}[tc][tc][1.0][0]{Square}
%\psfrag{a}{(a)}
%\psfrag{b}{(b)}
%\psfrag{aa}[tc][tl][1.0][0]{Triangular}
%\psfrag{ab}[tc][tl][0.9][0]{t=0.05}
%\psfrag{ba}[tc][tl][1.0][0]{Square}
%\psfrag{bb}[tc][tl][0.9][0]{t=0.15}
\includegraphics[width=0.45\textwidth,clip]{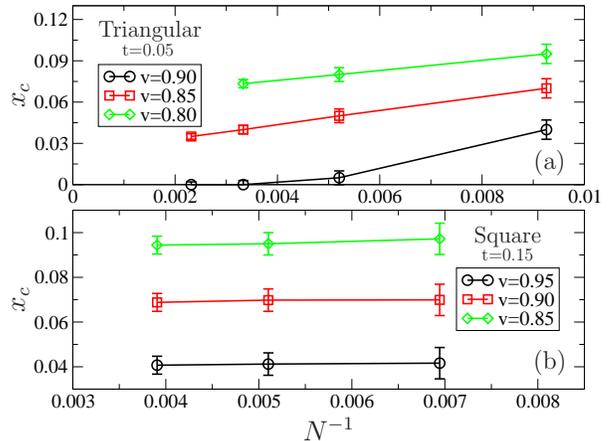}
\caption{\label{fig03}  (color online) Scaling of the critical doping $x_c$
defined by Maxwell construction with the inverse total number of sites. (a)
Triangular lattice. (b) Square lattice. }
\end{figure}

\begin{figure}[h]
\vspace{0.5cm}
%\psfrag{x}[tc][tc][1.1][0]{$x$}
%\psfrag{v}[tl][tc][1.1][0]{$v$}
%\psfrag{s}[tl][tc][1.0][0]{Square}
%\psfrag{t}[tl][tc][1.0][0]{Triangular}
\includegraphics[width=0.4\textwidth,clip]{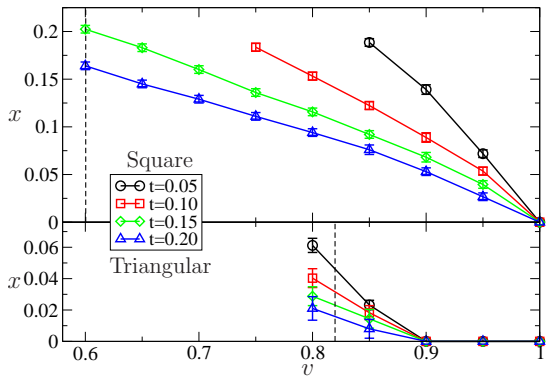}
\caption{\label{fig04} (color online) Phase separation boundaries for the
square and triangular lattices in the thermodynamic limit. The dashed lines
correspond to the approximate location of the phase transition between
plaquette and columnar phases \cite{syljuasen} for the square lattice and
between plaquette and RVB phases \cite{ralko,moessner} for the triangular
lattice.}
\end{figure}

To obtain the phase diagram in the $(v,x)$ plane, we have performed a systematic
size-scaling analysis at fixed $t$
and for various $v$'s depicted in Fig.\ref{fig03}.
% Figure Size-scaling {{{2
% }}}2
In agreement with the previous discussion, a significant size
dependence is only present for the triangular lattice, in which case PS
disappears for large clusters in the RVB phase in the vicinity
of the RK point \cite{ergodicity}.
In Fig.\ref{fig04}, we
report the thermodynamic limit of $x_c$ for the two lattices as a 
function of $v$, and for different values of $t$. For the square
lattice, calculations have been done from the RK point down to the expected phase
transition between
the plaquette phase and the columnar phase, namely $v \simeq 0.6$
\cite{syljuasen}. For the triangular lattice, the range between the RK point
down to the RVB-$\sqrt{12}\times\sqrt{12}$ transition point at $v \simeq
0.8$ has been covered \cite{ralko}.
% Figure Phase-Separation {{{2
% }}}2
These results clearly demonstrate the difference between the square and 
triangular lattices. In the first case, as soon as $v \neq 1$,
PS occurs for $x < x_c$. Moreover, upon decreasing $v$, crystalline order
strengthens and, for fixed $t$, it
is necessary to consider a higher concentration of holes to reach a stable
conducting phase. Similarly, the bigger $t$, the lower $x_c$.
 On the triangular
lattice, a finite size-scaling analysis shows that no phase
separation appears down to a critical value $v \sim 0.9$, well
above the critical value $v \sim 0.8$ below which plaquette order sets
in. Although numerical
limitations prevent computations for smaller $v$ and $t$, our
results up to the $3\times 12 \times 12$-site cluster provide clear
evidence for a region of PS inside the RVB region, between $v \sim 0.8$ and $v
\sim 0.9$ \cite{note01}.
 
\begin{figure}[h]
\vspace{0.5cm}
%\psfrag{corr}[bc][tc][0.9][0]{~~$D^2(\vec{k})$}
%\psfrag{g}[tc][tc][0.8][0]{$\Gamma$}
%\psfrag{m}[tc][tc][0.8][0]{$M$}
%\psfrag{k}[tc][tc][0.8][0]{$K$}
%\psfrag{kx}[tc][tc][0.8][0]{k$_x$}
%\psfrag{ky}[tc][tc][0.8][0]{k$_y$}
%\psfrag{aa}[tc][tc][1.0][0]{(c)}
%\psfrag{ab}[tc][tc][1.0][0]{(b)}
%\psfrag{ac}[tc][tc][1.0][0]{(a)}
%\psfrag{corrm}[bl][bl][0.9][0]{\hspace{-0.1cm}$D^{2} (\vec{k}_M)$}
%\psfrag{x}[tc][tc][1.1][0]{$x$}
%\psfrag{a}[tc][tc][0.9][0]{$\fbox{v=0.90, t=0.15}$}
\includegraphics[width=0.45\textwidth,clip]{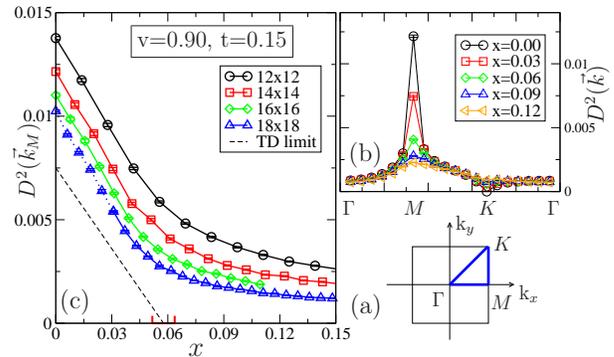}
\caption{\label{fig05} (color online) (a) First Brillouin zone. (b) Momentum dependence of the squared order parameter for the 196-site
cluster (14x14). (c) Squared order parameter at the $M$ point as a
function of the hole concentration $x$ and for different cluster sizes.
The thermodynamic limit is depicted as a dashed line, with the corresponding
error bar (see main text).}
\end{figure}

{\it Dimer ordering on the square lattice:} Next we
investigate how dimer order, known to exist at $x=0$, evolves under finite
doping. Two scenarii are {\it a priori} possible: i) the dimer order vanishes 
in the stable conducting phase
immediately at $x_c$; ii) dimer
order survives above $x_c$ in a narrow region of the conducting
phase. To solve this problem, we have calculated the squared order parameter 
$D^2 (\vec{k})$ in the GS $| \Psi_0 \rangle$ defined
by:
\begin{eqnarray}
D^2 (\vec{k}) = \frac{1}{N} \frac{\langle \Psi_0 |
d(-\vec{k}) d (\vec{k})| \Psi_0 \rangle}{\langle \Psi_0 | \Psi_0 \rangle}
\end{eqnarray}
along the path of Fig.\ref{fig05}(a) the
first Brillouin zone of the square lattice, where $d(\vec{k})$ is the Fourier
transform of the dimer operator defined on the horizontal bonds.
Note that this calculation has not been tried for the triangular lattice
since no Bragg peak is present in the RVB phase, and the algorithm is losing 
efficiency for $v \lesssim 0.8$~\cite{ergodicity}. In
the pure plaquette phase on the square lattice, a Bragg peak 
develops at point $\vec{k}_M=(\pi,0)$, the middle of the side of the BZ. We
show in Fig.\ref{fig05}(b) a typical result for the squared order parameter
on the $196$-site cluster for different values of $x$. Clearly, the Bragg peak
disappears upon doping.
% Figure Order parameter {{{2
% }}}2
A finite size scaling of the order parameter can be performed
thanks to the linear behaviour at low concentration. Within our data,
$D^2 (\vec{k} = \vec{k}_M) \equiv D^2_M (L,x)$ behaves like $a_L x + b_L$ in the linear region. In
this case, one can determine rather precisely $x_{+ \infty}$ such that $D^2_M
(+ \infty,x_{+ \infty})=0$ {\it i.e.} $x_{+ \infty}$ is the concentration in
the thermodynamic limit where the Bragg peak vanishes. By definition,
$x_{+ \infty} = - {b_{+ \infty}}/{ a_{+ \infty}} \simeq 0.05(8)$.
This value, and the linear behaviour in the thermodynamic limit, are
displayed in Fig.\ref{fig05}(c). If we compare $x_{+ \infty}$
to the corresponding $x_c$ from Fig.\ref{fig04}, we can
conclude that the Bragg peak indeed vanishes at $x_c \simeq
0.067(5)$ within error bars. Note that numerical errors
increase for larger clusters, and we are not able to use the
same analysis for clusters larger than
$256$ sites (the results for the $18 \times 18$ cluster have not been used). Although the determination of
$x_{+ \infty}$ is delicate, the linear behavior of $D^2$ vs $x$ is consistent
with the physical behavior expected for the binary system of
Fig.\ref{fig04}. No dimer order is present above $x_c$, showing that the
system is simply \textquotedblleft conducting \textquotedblright in this
case, with $D^2 (\vec{k}_{M})$ decreasing as $N^{-1}$ (critical
behaviour).

\begin{figure}[h]
\vspace{0.5cm}
\includegraphics[width=0.45\textwidth,clip]{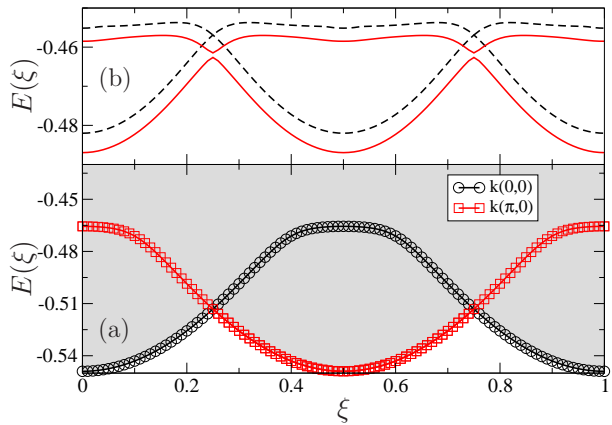}
\caption{\label{fig06} (color online) Energy spectrum vs 
(reduced) Aharonov-Bohm flux $\xi$, for a $4 \times 4$ cluster with
$4$ holes at $v=0.70$ and $t=1.00$. 
(a) Torus geometry. 
%The momenta ${\bf q}$ of the two low-energy branches
%are specified and the complete spectrum is displayed up
%to $\xi = 0.2$. 
Arbitrary BC are used in the transverse direction to obtain a
continuous set of momenta leading to a continuum (colored area).
(b) The two low-energy branches for the case of a
cylinder (dashed line) and including a small bond disorder (full red line). 
}
\end{figure}

{\it Flux quantization:}
Finally, let us turn to a better characterization of the
\textquotedblleft conducting phase\textquotedblright.
Since holes are bosonic one expects the conducting phase to be a 
superfluid (through Bose condensation)\cite{kivelson}.
However, extra complexity results from the fact that
these bosons move in a dimer environment.
To investigate this issue, we pierce the
torus by an Aharonov-Bohm flux of strength $\phi = \xi \phi_0$ with
 $0 \le \xi \le 1$ and where $\phi_0= h c / e$ is the magnetic flux quantum. 
To reduce finite size effects, we also consider 
arbitrary BC in the second direction (y).
All this is implemented by the Peierls substitution, 
changing the hole hopping term into 
$t' = \exp(\pm i 2 \pi \xi a / L_x\pm i 2 \pi \eta a / L_y) t$,
where the $\pm$ depends on the directions $\pm x\pm y$,
while $a$ is the lattice parameter and $L_x$ and $L_y$ the linear sizes 
of the system. Obviously,
the whole spectrum should be periodic in $\xi$ with period $1$.
We show in Fig.\ref{fig06} the
spectrum of the $4\times 4$ cluster on the square lattice, with $4$ holes.
% Figure Flux {{{2
% }}}2
It turns out that (i) the GS energy exhibits well-defined minima 
and (ii) is rigorously periodic with period $\xi=1/2$,
which means that there is flux quantization in units of {\it half} the
flux quantum (red curve). 
Property (i) is typical of a superfluid~\cite{Yang}: It is the precursor of a finite
barrier in the thermodynamic limit.~\cite{note_Yang}. 
Property (ii) was suggested quite some time ago
in the context of a more general
QDM by Kivelson\cite{kivelson}, who also predicted that, in the cylinder geometry, 
one should be able to tunnel between the two branches of Fig.\ref{fig06}, thus 
lifting the degeneracy at the level crossing.
This degeneracy is not lifted in our case, neither in the torus geometry,
nor in the cylinder geometry, due to the translational
symmetry, which puts the two states that are degenerate at $\xi=1/4$ into different symmetry 
sectors. However, getting rid of the 
translational symmetry by changing the amplitude of a local dimer flip indeed 
removes the degeneracy (upper panel of Fig.\ref{fig06}), leading to a 
detectable flux quantization in units of half the flux quantum
in an experiment in which the flux is sweeped.
Thus, in our model, the ground-state energy
has periodicity $h c / 2 e$, consistent with mobile elementary
particles of charge $Q=2e$ in the system.
Unlike what was recently found in a bosonic model with correlated hopping\cite{bendjama}, 
these particles are 
not boson pairs: 
%(although a definite proof would require
%the knowledge of the two particle Greens's function not accessible with GFMC,
%boson pairs are unlikely, 
%at least far enough from the phase separation: 
>From the bosonic
point of view, it is the statistical flux of the dimer background that leads to the half-flux 
quantization. If dimers are interpreted as SU(2) electron singlets, these singlets are the 
physical pairs that lead to half-flux quantization. This
scenario is fundamentally different from the usual mechanism related to real
space pairing of the charge carriers found e.g. in the extended Hubbard chain\cite{penc},
in the 2-leg ladders \cite{hayward} or, more generally, in Luther-Emery liquids
\cite{seidel}, as can be inferred from the exact degeneracy between $\xi=0$ and
$\xi=0.5$ for finite systems in the present case, to be contrasted with the significant 
finite-size effects of the other cases.

{\it Summary and conclusions:} The numerical investigation with Green's function
Quantum Monte Carlo and exact diagonalizations of the doped two-dimensional quantum hard-core 
dimer model on the square and triangular
lattices has led to a number of interesting conclusions regarding hole motion in
a dimer background. 
Phase separation is often present at low doping, as suggested by
earlier investigations, but our results indicate that it is related to the 
presence of valence bond order \cite{syljuasen2}: 
In the RVB phase
of the triangular lattice, PS only occurs close to the plaquette phase, where short-range
dimer correlations are already strong enough. Close to the RK point, 
doping the RVB phase leads
directly to a superfluid phase as shown from its response to
an Aharonov-Bohm flux. Moreover, we observed that the flux quantization is in units of 
half a flux quantum,
consistent with the idea that the dimer background leads to effective particles
of charge $2e$.  All these results are in qualitative agreement with the gauge
theories of high T$_c$ superconductivity in strongly correlated systems~\cite{PALee}.

% Acknowledgements           {{{1
\begin{acknowledgments}
We acknowledge useful discussions with Federico Becca. This work was supported
 by the Swiss National Fund, by MaNEP, and by the Agence Nationale de la Recherche
(France).
\end{acknowledgments}
% }}}1
% Bibliography               {{{1


\begin{thebibliography}{99}
\bibitem{anderson} P.W. Anderson, Science {\bf 235}, 1196 (1987).
\bibitem{rokhsar} D.S. Rokhsar and S.A. Kivelson, \prl {\bf 61}, 2376 (1988).
\bibitem{vernay} F. Vernay, A. Ralko, F. Becca and F. Mila, \prb {\bf 74},
054402 (2006).
\bibitem{zhitomirsky} M. E. Zhitomirsky, \prb {\bf 71}, 214413 (2005).
\bibitem{poilblanc} D. Poilblanc, F. Alet, F. Becca, A. Ralko, F. Trousselet
and F. Mila, \prb {\bf 74}, 014437 (2006).
\bibitem{Castelnovo} C. Castelnovo, C. Chamon, C. Mudry and P. Pujol, Ann.
Phys. {\bf 322}, 903 (2007).
\bibitem{nandini} N. Trivedi and D.M. Ceperley, \prb {\bf 41}, 4552 (1990); M. Calandra and S. Sorella, \prb {\bf 57}, 11446 (1998).
\bibitem{ralko} A. Ralko, M. Ferrero, F. Becca, D. Ivanov, and F. Mila, \prb {\bf 74}, 134301 (2006)
and references therein.
\bibitem{ergodicity} For small values of $t$, the GFMC algorithm is no longer
ergodic due to hole localization.
\bibitem{syljuasen} O.F. Syljuasen, \prb {\bf 73}, 245105 (2006).
\bibitem{note01} 
Long-range Coulomb repulsion is expected to have a
major role in the PS region and might stabilize stripes.
\bibitem{moessner} R. Moessner and S. L. Sondhi, \prb {\bf 63}, 224401 (2001).
\bibitem{kivelson} S. Kivelson, \prb {\bf 39}, 259 (1989).
\bibitem{Yang} N.~Byers and C.N.~Yang, \prl~{\bf 7}, 46 (1961). 
\bibitem{note_Yang} By contrast, for non-interacting fermions, signs of a flat energy curve 
already appear on such small clusters
provided one also uses arbitrary BC. 
Similar arguments were 
used in D.~Poilblanc, \prb~{\bf 44}, 9562 (1991) for the 2D t-J model.
\bibitem{bendjama} R. Bendjama, B. Kumar, F. Mila, \prl {\bf 95},
  110406 (2005).
\bibitem{penc} K. Penc and F. Mila, \prb {\bf 49}, 9670 (1994).
\bibitem{hayward} C. A. Hayward,
D. Poilblanc, R. M. Noack, D. J. Scalapino and W. Hanke, \prl {\bf 75}, 926
(1995).
\bibitem{seidel} A. Seidel and D. H. Lee, \prb {\bf 71},
045113 (2005).
\bibitem{syljuasen2} Our results give an explicit
characterization of the confinement-deconfinement discussed in O.F. Syljuasen,
\prb {\bf 71}, 020401(R)(2005).
\bibitem{PALee} T. Senthil and P.A.~Lee, \prb {\bf 71}, 174515 (2005).
  and references therein.
\end{thebibliography}
\end{document}